\documentclass[a4paper]{article}

\usepackage[margin=1in]{geometry}

\usepackage[T1]{fontenc}
\newcommand{\changefont}[3]{
\fontfamily{#1} \fontseries{#2} \fontshape{#3} \selectfont}

\changefont{ptm}{m}{n}

\usepackage{setspace} 
\usepackage{graphicx}    

\usepackage{epstopdf}
\usepackage{amsfonts}
\usepackage{amssymb}
\usepackage{graphics}
\usepackage{mathrsfs}
\usepackage{color}
\newtheorem{remark}{Remark}[section]

\newtheorem{theorem}{Theorem}[section]

\long\def\symbolfootnote[#1]#2{\begingroup%
\def\thefootnote{\fnsymbol{footnote}}\footnote[#1]{#2}\endgroup} 

\begin{document}

\begin{center}
\Large \textbf{Period-doubling route to chaos in driven impulsive systems}
\end{center}

\begin{center}
\normalsize \textbf{Mehmet Onur Fen$^1$, Fatma Tokmak Fen$^2$} \\
\vspace{0.2cm}
\textit{\textbf{\footnotesize $^1$Department of Mathematics, TED University, 06420 Ankara, Turkey}}\\ 
\vspace{0.1cm}
\textit{\textbf{\footnotesize $^2$Department of Mathematics, Gazi University, 06500 Ankara, Turkey}} 
\vspace{0.1cm}
\end{center}

\vspace{0.3cm}

\begin{center}
\textbf{Abstract}
\end{center}

\noindent\ignorespaces

In the present study, we investigate the dynamics of impulsive differential equations driven by a chaotic system. We rigorously prove that, likewise the drive, the response impulsive system is also chaotic. Our results are based on the presence of sensitivity and infinitely many unstable periodic motions. The theoretical results are supported by simulations. 


\vspace{0.2cm}
 
\noindent\ignorespaces \textbf{Keywords:} Impulsive systems, Sensitivity, Period-doubling cascade, Unidirectional coupling.

\vspace{0.6cm}


\section{Introduction}

One of the routes to chaos is the period-doubling cascade, which was first observed in quadratic maps \cite{Myrberg}. This phenomenon is based on the successive emergence of periodic motions with twice period of the previous oscillation as some parameter is varied in a system \cite{Moon04}-\cite{Sander11}. A critical parameter value exists at which the process accumulates, and beyond the critical value the system possesses chaotic motions \cite{Moon04}. The period-doubling onset of chaos exhibits a universal behavior \cite{Feigenbaum80}. Period-doubling route to chaos can be observed in various fields of science such as mechanics, electrical circuits, lasers, magnetism, photochemistry, neural processes, predator-prey systems, and glow discharge-semiconductor systems \cite{Letellier07}-\cite{Pourazarm15}. 

In this study, we investigate the formation of period-doubling cascade in driven impulsive systems. Impulsive differential equations describe the dynamics of real world processes in which abrupt changes occur. Such equations play an increasingly important role in various fields such as mechanics, electronics, biology, neural networks, communication systems, chaos theory, and population dynamics \cite{Akh1}-\cite{Yang97b}. In the present paper, we take into account unidirectionally coupled systems such that the drive system is chaotic through period-doubling cascade and the response system admits impulsive actions. We rigorously prove that the response also possesses chaos through period-doubling cascade under sufficient conditions. Our results are based on the presence of sensitivity, which is the main ingredient of chaos \cite{Lorenz63}-\cite{Rob95}, and infinitely many unstable periodic solutions.

Chaos in impulsive systems was also considered in the study \cite{Akhmet2014b}. The presence of chaos in the sense of Li-Yorke was investigated in \cite{Akhmet2014b} such that the proximality and frequent separation features were theoretically proved. Differenly from the paper \cite{Akhmet2014b}, in the present study, we investigate sensitivity and period-doubling route to chaos in impulsive systems. Illustrative examples that support the theoretical results are also provided.

The rest of the paper is organized as follows. In Section \ref{sec_model}, we introduce the systems of differential equations that will be investigated. Sufficient conditions for the formation of period-doubling route to chaos in impulsive systems as well as bounded solutions are considered in Section \ref{sec_bounded}. In Section \ref{sec_sensitivity}, the presence of sensitivity in impulsive systems is rigorously proved. Section \ref{imp_pdc} is devoted to the formation of period-doubling cascades, and finally, examples that support the theoretical results are provided in Section \ref{sec_examples}.

\section{The model} \label{sec_model}

Let us consider the system 
\begin{eqnarray} \label{main_part}
x'= F(t,x), 
\end{eqnarray}
where $F:\mathbb R \times \mathbb R^m \to \mathbb R^m$ is a continuous function that is periodic in $t$, i.e., there exists a positive number $T$ such that $F(t+T,x)=F(t,x)$ for all $t \in \mathbb R$ and $x\in \mathbb R^m$. Our main assumption on system (\ref{main_part}) is the existence of a nonempty set $\mathscr{A}$ of all solutions of (\ref{main_part}) that are uniformly bounded on $\mathbb R.$ In this case, there exists a compact set $\Lambda \subset \mathbb R^m$ such that the trajectories of all solutions that belong to $\mathscr{A}$ lie inside $\Lambda$.

Next, we take into account the impulsive system
\begin{eqnarray}
\begin{array}{l} \label{impulsive_part}
y' = A y + f(t,y) + g(x(t)),  ~t \neq  \theta_k, \\
\Delta y |_{t = \theta_k} = B y + W(y),
\end{array}
\end{eqnarray}
where $x(t)$ is a solution of (\ref{main_part}), the functions $f:\mathbb R\times \mathbb R^n\to \mathbb R^n,$ $g: \mathbb R^m \to \mathbb R^n$ and $W:\mathbb R^n \to \mathbb R^n$ are continuous in all their arguments, the function $f(t,y)$ satisfies the periodicity condition $f(t+T,y)=f(t,y)$ for all $t\in\mathbb R$, $y\in \mathbb R^n$, $A$ and $B$ are constant, $n \times n$ real matrices, the sequence $\left\{\theta_k\right\},$ $k\in \mathbb Z,$ of impulsive moments is strictly increasing, $\Delta y |_{t = \theta_k}=y(\theta_k+)-y(\theta_k)$, and $y(\theta_k + )=\displaystyle \lim_{t\to \theta_k^+} y(t).$ We suppose that there exists a natural number $p$ such that $\theta_{k+p}=\theta_k+T$ for all $k\in\mathbb Z$.

We will rigorously prove that if system (\ref{main_part}) is chaotic through period-doubling cascade, then the same is true for the impulsive system (\ref{impulsive_part}). The description of period-doubling cascade for systems (\ref{main_part}) and (\ref{impulsive_part}) will be provided in Section \ref{imp_pdc}.

\section{Bounded solutions} \label{sec_bounded}

Throughout the paper, we will make use of the usual Euclidean norm for vectors and the norm induced by the Euclidean norm for square matrices.

The following conditions are required.
\begin{enumerate}
\item[\bf (A1)] The matrices $A$ and $B$ commute, and $\det(I+B)\neq 0,$ where $I$ is the $n\times n$ identity matrix;

\item[\bf (A2)] The eigenvalues of the matrix $A+\displaystyle \frac{p}{T} \ln (I+B)$ have negative real parts;

\item[\bf (A3)] There exist positive numbers $M_f$ and $M_W$ such that $\displaystyle \sup_{t\in \mathbb R, y\in \mathbb R^n} \left\|f(t,y)\right\|\le M_f$ and $\displaystyle \sup_{y\in \mathbb R^n} \left\|W(y)\right\|\le M_W$.

\item[\bf (A4)] There exist positive numbers $L_F$, $L_f$, $L_1$, $L_2$, and $L_W$ such that 

\textbf{(i)} $\left\|F(t,x_1)-F(t,x_2)\right\| \leq L_F\left\|x_1-x_2\right\|$ for all $t\in\mathbb R,$ $x_1, x_2 \in \Lambda,$ 

\textbf{(ii)} $\left\|f(t,y_1)-f(t,y_2)\right\| \leq L_f\left\|y_1-y_2\right\|$ for all $t\in\mathbb R,$ $y_1, y_2 \in \mathbb R^{n},$

\textbf{(iii)} $L_1\left\|x_1-x_2\right\| \le \left\|g(x_1)-g(x_2)\right\| \leq L_2\left\|x_1-x_2\right\|$ for all $x_1, x_2\in\Lambda,$

\textbf{(iv)} $\left\|W(y_1)-W(y_2)\right\| \leq L_W\left\|y_1-y_2\right\|$ for all $y_1,y_2 \in \mathbb R^{n}.$ 

\end{enumerate}

Let us denote by $U(t,s)$ the transition matrix of the linear homogeneous impulsive system
\begin{eqnarray*}
&& u' = A u,   ~t \neq  \theta_k, \\
&& \Delta u |_{t = \theta_k} = B u.
\end{eqnarray*}
Under the conditions $(A1)$ and $(A2)$, there exist positive numbers $N$ and $\omega$ such that $\left\|U(t,s)\right\| \le Ne^{-\omega (t-s)}$ for $t\ge s$ \cite{Akh1,Samolienko95}. 

The following conditions are also needed. 
\begin{enumerate}
\item[\bf (A5)] $\displaystyle N \left( \frac{L_f}{\omega} + \frac{pL_W}{1-e^{-\omega T}}  \right) <1;$

\item[\bf (A6)] $\displaystyle -\omega+NL_f+\frac{p}{T} \ln (1+NL_W)<0;$

\item[\bf (A7)] $L_W \left\| \left(I+B\right)^{-1}  \right\|<1.$
\end{enumerate} 

For a fixed solution $x(t)$ of (\ref{main_part}), a left-continuous function $y(t) : \mathbb R \to \mathbb R^n$ is a solution of (\ref{impulsive_part}) if: (i) It has discontinuities only at the points $\theta_k,$ $k\in \mathbb Z,$ and these discontinuities are of the first kind; (ii) The derivative $y'(t)$ exists at each point $t \in \mathbb R \setminus \left\{\theta_k\right\}$, and the left-sided derivative exists at the points $\theta_k,$ $k\in\mathbb Z;$ (iii) The differential equation is satisfied by $y(t)$ on $\mathbb R \setminus \left\{\theta_k\right\},$ and it holds for the left derivative of $y(t)$ at every point $\theta_k$, $k\in\mathbb Z;$ (iv) The jump equation is satisfied by $y(t)$ for every $k\in\mathbb Z.$

According to the results of \cite{Akh1,Samolienko95},  under the conditions $(A1),(A2),$ $(A3), (ii), (iv)$ $(A4), (ii), (iv)$, and $(A5)$, for each $x(t) \in \mathscr{A}$ there exists a unique solution $\phi_{x(t)}(t)$ of system (\ref{impulsive_part}) that is bounded on the whole real axis, and it satisfies the relation
\begin{eqnarray*} 
\phi_{x(t)}(t)=\displaystyle \int_{-\infty}^t U(t,s) \left[ f(s,\phi_{x(t)}(s)) + g(x(s))\right] ds + \sum_{-\infty< \theta_k<t} U(t,\theta_k+) W (\phi_{x(t)}(\theta_k)).
\end{eqnarray*}
Let us denote $$M_g=\displaystyle \sup_{x\in\Lambda} \left\|g(x)\right\|.$$ It can be verified that the inequality $\displaystyle \sup_{t \in \mathbb R} \left\|\phi_{x(t)}(t)\right\| \leq K_0$ is valid for each $x(t) \in \mathscr{A}$, where $$K_0=\displaystyle \frac{N (M_f+M_g)}{\omega} + \frac{pNM_W}{1-e^{-\omega T}}.$$  
Moreover, if condition $(A6)$ additionally holds, then for a fixed solution $x(t) \in \mathscr{A}$ of (\ref{main_part}) the bounded solution $\phi_{x(t)}(t)$  attracts all other solutions of (\ref{impulsive_part}). That is, $\left\| y(t) - \phi_{x(t)}(t) \right\| \to 0$ as $t\to \infty$ for each solution $y(t)$ of (\ref{impulsive_part}).

It is worth noting that the results of the present paper are valid even if we replace the non-autonomous system (\ref{main_part}) by an autonomous one of the form
\begin{eqnarray*}  
x'= \overline{F}(x) 
\end{eqnarray*} 
with the counterpart of condition $(A4), (i)$, where $\overline{F}: \mathbb R^m \to \mathbb R^m$ is a continuous function. 

In the next section the presence of sensitivity in the impulsive system (\ref{impulsive_part}) will be considered.

\section{Sensitivity analysis} \label{sec_sensitivity}

System (\ref{main_part}) is called sensitive if there exist positive numbers $\epsilon_0$ and $\Delta$ such that for an arbitrary positive number $\delta_0$ and for each $x(t) \in \mathscr{A},$ there exist $\overline{x}(t) \in \mathscr{A},$ $t_0 \in \mathbb R$ and an interval $J \subset [t_0, \infty),$ with a length no less than $\Delta,$ such that $\left\|x(t_0)-\overline{x}(t_0)\right\|<\delta_0$ and $\left\|x(t)-\overline{x}(t)\right\| > \epsilon_0,$ $t \in J$ \cite{Akhmet2013}.

We say that system (\ref{impulsive_part}) replicates the sensitivity of (\ref{main_part}) if there exist positive numbers $\epsilon_1$ and $\overline{\Delta}$ such that for an arbitrary positive number $\delta_1$ and for each bounded solution $\phi_{x(t)}(t)$ of (\ref{impulsive_part}), there exist a bounded solution $\phi_{\overline{x}(t)}(t)$ of the same system, $t_0\in\mathbb R$ and an interval $J^1\subset [t_0,\infty),$ with a length no less than $\overline{\Delta},$ such that the inequalities $\left\|\phi_{x(t)}(t_0)-\phi_{\overline{x}(t)}(t_0)\right\|<\delta_1$ and $\left\|\phi_{x(t)}(t)-\phi_{\overline{x}(t)}(t)\right\| > \epsilon_1,$ $t \in J^1,$ hold, and $J^1$ contains at most one element of the sequence $\left\{\theta_k\right\}.$

In what follows, we will denote by $i(\Gamma)$ the number of the terms of the sequence $\left\{\theta_k\right\}$, $k\in \mathbb Z$, which belong to an interval $\Gamma$. One can confirm that $\displaystyle i((a,b))\leq p+\frac{p}{T}(b-a),$ where $a$ and $b$ are numbers such that $b>a$.

\begin{theorem}\label{imp_pdc_sensitivity_thm}
Under the conditions $(A1)-(A7),$ the impulsive system (\ref{impulsive_part}) replicates the sensitivity of (\ref{main_part}).
\end{theorem}

\noindent \textbf{Proof.}
Fix an arbitrary number $\delta_1>0$ and a bounded solution $\phi_{x(t)}(t)$ of (\ref{impulsive_part}), where $x(t)\in \mathscr{A}.$ Let $\displaystyle \alpha= \omega - NL_f - \frac{p}{T} \ln (1+NL_W).$ Note that the number $\alpha$ is positive by condition $(A6).$ Suppose that $\overline{\epsilon}$ is a sufficiently small positive number such that 
$$
\left[ 1+\frac{NL_2}{\omega} \left(  1+ \frac{NL_f}{\alpha} (1+NL_W)^p + \frac{pNL_W }{1-e^{-\alpha T}} (1+NL_W)^p  \right)    \right] \overline{\epsilon} \le \delta_1.
$$ 
Take a number $R<0$ sufficiently large in absolute value such that 
$$
\left( \frac{2N(M_f+M_g)}{\omega} + \frac{2pNM_W }{1-e^{-\omega T}} \right) (1+NL_W)^p e^{\alpha R} \le \overline{\epsilon},
$$ 
and let $\delta_0=\overline{\epsilon}e^{L_F R}.$

Since (\ref{main_part}) is sensitive, there exist positive numbers $\epsilon_0$ and $\Delta$ such that $\left\|x(t_0)-\overline{x}(t_0)\right\|<\delta_0$ and $\left\|x(t)-\overline{x}(t)\right\| > \epsilon_0,$ $t \in J,$ for some $\overline{x}(t)\in \mathscr{A},$ $t_0 \in \mathbb R$ and for some interval $J \subset [t_0,\infty)$ whose length is not less than $\Delta.$
In the first part of the proof, we will show that $\left\|\phi_{x(t)}(t_0)-\phi_{\overline{x}(t)}(t_0)\right\|<\delta_1.$ 

The solutions $x(t)$ and $\overline{x}(t)$ satisfy the integral equation
$$
x(t)-\overline{x}(t)=x(t_0)-\overline{x}(t_0)+\displaystyle \int_{t_0}^t [F \left(s,x(s)\right) -  F\left(s,\overline{x}(s)\right) ] ds.
$$
Therefore, we have for $t\in [t_0+R, t_0]$ that 
$$
\left\|x(t)-\overline{x}(t)\right\| \le \left\|x(t_0)-\overline{x}(t_0)\right\| + \displaystyle \left| \int^t_{t_0}  L_F \left\|x(s)-\overline{x}(s)\right\|  ds \right|.
$$
By means of the Gronwall-Bellman inequality, one can confirm that
\begin{eqnarray*}
\left\|x(t)-\overline{x}(t)\right\| \leq \left\|x(t_0)-\overline{x}(t_0)\right\|e^{L_F\left|t-t_0\right|}.
\end{eqnarray*} 
Hence, $\left\|x(t)-\overline{x}(t)\right\|<\overline{\epsilon}$ for $t \in [t_0+R, t_0].$ 

Since the relation
\begin{eqnarray*}
&& \phi_{x(t)}(t) - \phi_{\overline{x}(t)}(t) = \displaystyle \int_{-\infty}^{t_0+R} U(t,s) \left[  f\left(s,\phi_{x(t)}(s)\right) + g(x(s)) - f\left(s,\phi_{\overline{x}(t)}(s)\right) - g( \overline{x}(s) ) \right] ds \\
&& + \displaystyle \int^{t}_{t_0+R} U(t,s) \left[ f\left(s,\phi_{x(t)}(s)\right) - f\left(s,\phi_{\overline{x}(t)}(s)\right) \right] ds  + \displaystyle \int^{t}_{t_0+R} U(t,s) \left[ g(x(s)) - g(\overline{x}(s)) \right] ds \\
&& + \displaystyle \sum_{-\infty < \theta_k \le t_0+R} U(t,\theta_k+) \left[  W\left( \phi_{x(t)}(\theta_k) \right) - W \left( \phi_{\overline{x}(t)}(\theta_k) \right)  \right] \\
&& + \displaystyle \sum_{t_0+R < \theta_k < t} U(t,\theta_k+) \left[  W \left(  \phi_{x(t)}(\theta_k) \right)  - W \left( \phi_{\overline{x}(t)}(\theta_k) \right)  \right]
\end{eqnarray*}
holds, we have that
\begin{eqnarray} \label{imp_pdc_proof1}
\begin{array}{l}
\left\| \phi_{x(t)}(t) - \phi_{\overline{x}(t)}(t) \right\| \le \left(\displaystyle  \frac{2N(M_f+M_g)}{\omega} +  \frac{2pNM_W}{1-e^{-\omega T}}  \right)e^{-\omega (t-t_0-R)} + \displaystyle \frac{NL_2 \overline{\epsilon}}{\omega} \left(  1- e^{-\omega(t-t_0-R)}  \right)   \\
 + \displaystyle \int^t_{t_0+R} NL_f e^{-\omega (t-s)} \left\| \phi_{x(t)}(s) - \phi_{\overline{x}(t)}(s) \right\| ds  +  \displaystyle \sum_{t_0+R < \theta_k < t} NL_W e^{-\omega (t-\theta_k)} \left\| \phi_{x(t)}(\theta_k) - \phi_{\overline{x}(t)}(\theta_k) \right\|.
\end{array}
\end{eqnarray}

Let us define the functions $\nu(t)=e^{\omega t} \left\| \phi_{x(t)}(t) - \phi_{\overline{x}(t)}(t) \right\|$ and $h(t)=c+\displaystyle \frac{NL_2 \overline{\epsilon}}{\omega}e^{\omega t},$ where $$c=\left(\displaystyle  \frac{2N(M_f+M_g)-NL_2 \overline{\epsilon}}{\omega} + \frac{2pNM_W}{1-e^{-\omega T}} \right) e^{\omega (t_0+R)}.$$ The inequality (\ref{imp_pdc_proof1}) implies that
$$
\nu(t) \le h(t) + \displaystyle \int^t_{t_0+R} NL_f \nu(s) ds + \sum_{t_0+R < \theta_k < t} NL_W \nu(\theta_k), ~ t\in [t_0+R,t_0].
$$
By applying the analogue of the Gronwall's inequality for piecewise continuous functions one can verify that
\begin{eqnarray*}
&& \nu(t) \le h(t) +  \displaystyle \int^t_{t_0+R} NL_f (1+NL_W)^{i((s,t))} e^{NL_f (t-s)} h(s) ds \\
&& + \sum_{t_0+R < \theta_k < t} NL_W  (1+NL_W)^{i((\theta_k,t))} e^{NL_f (t-\theta_k)} h(\theta_k).
\end{eqnarray*} 

Using the equation
\begin{eqnarray*}
&& 1+ \displaystyle \int_{t_0+R}^t NL_f (1+NL_W)^{i((s,t))} e^{NL_f (t-s)} ds + \displaystyle \sum_{t_0+R < \theta_k <t} NL_W (1+NL_W)^{i((\theta_k,t))} e^{NL_f (t-\theta_k)} \\
&& = (1+ NL_W)^{i((t_0+R,t))} e^{NL_f(t-t_0-R)} 
\end{eqnarray*} 
together with the inequality
$$
(1+NL_W)^{i((a,b))} e^{NL_f(b-a)} \le (1+NL_W)^p e^{(\omega-\alpha)(b-a)}, ~b\ge a,
$$
we obtain that
\begin{eqnarray*}
&& \nu(t) \le c(1+NL_W)^p e^{(\omega-\alpha)(t-t_0-R)} + \frac{NL_2\overline{\epsilon}}{\omega} e^{\omega t} \\
&& + \displaystyle \int_{t_0+R}^t \frac{N^2 L_f L_2  \overline{\epsilon}}{\omega} (1+NL_W)^p e^{(\omega-\alpha)(t-s)} e^{\omega s} ds \\
&& + \displaystyle \sum_{t_0+R < \theta_k < t} \frac{N^2 L_2 L_W  \overline{\epsilon}}{\omega} (1+NL_W)^p e^{(\omega-\alpha)(t-\theta_k)} e^{\omega \theta_k}. 
\end{eqnarray*} 
The last inequality implies for $t\in [t_0+R,t_0]$ that
\begin{eqnarray*}
&& \left\| \phi_{x(t)}(t) - \phi_{\overline{x}(t)}(t) \right\| \le \left(\displaystyle  \frac{2N(M_f+M_g)-NL_2 \overline{\epsilon}}{\omega} + \frac{2pNM_W}{1-e^{-\omega T}} \right) (1+NL_W)^p e^{-\alpha (t-t_0-R)} + \frac{NL_2\overline{\epsilon}}{\omega} \\
&& + \displaystyle \frac{N^2 L_f L_2 \overline{\epsilon}}{\alpha \omega} (1+NL_W)^p \left( 1-e^{-\alpha (t-t_0-R)} \right) 
   + \displaystyle \frac{pN^2 L_2 L_W \overline{\epsilon}}{(1-e^{-\alpha T}) \omega} (1+NL_W)^p \left( 1-e^{-\alpha (t-t_0-R+T)} \right).
\end{eqnarray*} 
Hence,
\begin{eqnarray*}
&& \left\| \phi_{x(t)}(t_0) - \phi_{\overline{x}(t)}(t_0) \right\| <  \left(\displaystyle  \frac{2N(M_f+M_g)}{\omega} + \frac{2pNM_W}{1-e^{-\omega T}} \right) (1+NL_W)^p e^{\alpha R} \\
&& + \displaystyle \frac{NL_2 \overline{\epsilon}}{\omega} \left( 1+ \frac{NL_f}{\alpha}  (1+NL_W)^p + \frac{p N L_W}{1-e^{-\alpha T}} (1+NL_W)^p \right) \\
&& \le \left[ 1+\frac{NL_2}{\omega} \left(  1+ \frac{NL_f}{\alpha} (1+NL_W)^p + \frac{pNL_W }{1-e^{-\alpha T}} (1+NL_W)^p  \right)    \right] \overline{\epsilon} \\
&& \le \delta_1.
\end{eqnarray*} 

Next, we will show the existence of positive numbers $\epsilon_1$ and $\overline{\Delta}$ such that $\left\|\phi_{x(t)}(t)-\phi_{\overline{x}(t)}(t)\right\| > \epsilon_1$ for each $t \in J^1,$ where $J^1\subset [t_0,\infty)$ is an interval which has a length $\overline{\Delta}$ and contains at most one element of the sequence $\left\{\theta_k\right\},$ $k \in \mathbb Z$, of impulsive moments.

Let us denote $M_F= \displaystyle \sup_{t\in \mathbb R, x\in\Lambda} \left\|F(t,x)\right\|$.
Since for each $x(t)\in \mathscr{A}$ the inequality $\displaystyle\sup_{t\in\mathbb R} \left\|x'(t)\right\| \le M_F$ holds, one can conclude that the set $\mathscr{A}$ is an equicontinuous family of functions on $\mathbb R.$  Suppose that $g(x)= \left(g_1(x), g_2(x), \cdots, g_n(x) \right),$ where each $g_j,$ $1 \leq j \leq n,$ is a real valued function. Because the function $\overline{g} : \Lambda \times \Lambda \to \mathbb R^n$ defined as $\overline{g}(x_1, x_2) = g(x_1)-g(x_2)$ is uniformly continuous on $\Lambda \times \Lambda$, the set consisting of the elements of the form $g_i(x(t)) - g_i(\overline{x}(t)),$ $i=1,2,\ldots,n,$ where $x(t),$ $\overline{x}(t)\in \mathscr{A},$ is an equicontinuous family on $\mathbb R.$ Therefore, there exists a positive number $\tau<\Delta,$ which does not depend on the functions $x(t)$ and $\overline{x}(t),$  such that for each $t_1,t_2\in \mathbb R$ with $\left|t_1-t_2\right|<\tau,$ the inequality 
\begin{eqnarray} \label{impulsive_pdc_proof1}
\left| \left(g_i\left(x(t_1)\right) - g_i\left(\overline{x}(t_1)\right)  \right) - \left(g_i\left(x(t_2)\right) - g_i\left(\overline{x}(t_2)\right)  \right)   \right| <\displaystyle \frac{L_1\epsilon_0}{2n}
\end{eqnarray}
is valid for all $i=1,2,\ldots,n.$

Let $\eta$ be the midpoint of the interval $J$ and $\zeta=\eta-\tau/2.$ There exists an integer $j,$ $1\le j \le n,$ such that
$$
\left| g_j(x(\eta)) - g_j(\overline{x}(\eta)) \right| \ge \frac{1}{n} \left\| g(x(\eta)) - g(\overline{x}(\eta)) \right\|,
$$
and therefore, condition $(A4),$ $(iii),$ implies that 
$$
\left| g_j(x(\eta)) - g_j(\overline{x}(\eta)) \right| \ge \frac{L_1}{n} \left\|x(\eta) - \overline{x} (\eta)\right\| > \frac{L_1\epsilon_0}{n}.
$$
According to (\ref{impulsive_pdc_proof1}), we have for all $t \in [\zeta, \zeta +\tau]$ that 
$$
\left| g_j(x(t)) - g_j(\overline{x}(t)) \right| > \left| g_j(x(\eta)) - g_j(\overline{x}(\eta)) \right| -  \frac{L_1 \epsilon_0}{2n} > \frac{L_1 \epsilon_0}{2n}.
$$
One can confirm by using the last inequality that 
$$
\Big\| \displaystyle \int_{\zeta}^{\zeta+\tau} [g(x(s)) - g(\overline{x}(s))] ds \Big\| > \frac{\tau L_1 \epsilon_0}{2n}.
$$
 
For $t\in [\zeta, \zeta+\tau],$ the functions $\phi_{x(t)}(t)$ and $\phi_{\overline{x}(t)}(t)$ satisfy the relations
\begin{eqnarray*}
&& \phi_{x(t)}(t) = \phi_{x(t)}(\zeta) + \displaystyle \int^t_{\zeta}  \left[ A  \phi_{x(t)}(s) + f\left(s,\phi_{x(t)}(s)\right) +g(x(s)) \right] ds \\
&& + \displaystyle \sum_{\zeta \le \theta_k < t} \left[ B \phi_{x(t)}(\theta_k) + W \left(\phi_{x(t)}(\theta_k)\right)\right]
\end{eqnarray*}
and
\begin{eqnarray*}
&& \phi_{\overline{x}(t)}(t) = \phi_{\overline{x}(t)}(\zeta) + \displaystyle \int^t_{\zeta}  \left[ A  \phi_{\overline{x}(t)}(s) + f\left(s,\phi_{\overline{x}(t)}(s)\right) +g(\overline{x}(s)) \right] ds \\
&& + \displaystyle \sum_{\zeta \le \theta_k < t} \left[ B \phi_{\overline{x}(t)}(\theta_k) + W \left( \phi_{\overline{x}(t)}(\theta_k) \right)  \right],
\end{eqnarray*}
respectively. Thus, we have that
\begin{eqnarray*}
&& \left\| \phi_{x(t)}(\zeta+\tau) - \phi_{\overline{x}(t)}(\zeta+\tau) \right\| \ge  \Big\| \displaystyle \int_{\zeta}^{\zeta+\tau} [g(x(s)) - g(\overline{x}(s))] ds \Big\| - \left\| \phi_{x(t)}(\zeta) - \phi_{\overline{x}(t)}(\zeta) \right\| \\
&& - \displaystyle \int_{\zeta}^{\zeta+\tau} \left( \left\| A \right\| + L_f \right) \left\| \phi_{x(t)}(s) - \phi_{\overline{x}(t)}(s) \right\| ds - \displaystyle \sum_{\zeta \le \theta_k < \zeta + \tau} \left(\left\|B\right\|+L_W\right) \left\| \phi_{x(t)}(\theta_k) - \phi_{\overline{x}(t)}(\theta_k) \right\| \\
&& > \frac{\tau L_1 \epsilon_0}{2n} - \left[ 1+ \tau (\left\|A\right\| + L_f) + p \left(1+\frac{\tau}{T}\right) (\left\|B\right\|+L_W)  \right] \sup_{t\in[\zeta,\zeta+\tau]} \left\| \phi_{x(t)}(t) - \phi_{\overline{x}(t)}(t) \right\|. 
\end{eqnarray*}

The last inequality implies that $\displaystyle \sup_{t\in[\zeta,\zeta+\tau]} \left\| \phi_{x(t)}(t) - \phi_{\overline{x}(t)}(t) \right\| > \overline{M},$ where
$$
\displaystyle \overline{M} = \frac{\tau L_1 \epsilon_0}{2n\left[ 2+ \tau (\left\|A\right\| + L_f) + p \left(1+\displaystyle \frac{\tau}{T}\right) (\left\|B\right\|+L_W)  \right]}.
$$
Set $\underline{\theta} = \displaystyle \min_{1 \le k \le p} \left( \theta_{k+1} - \theta_k \right),$ and define the numbers
$$
\epsilon_1= \displaystyle \frac{\overline{M}}{2} \min \displaystyle \left\{1, \frac{1-L_W \left\|(I+B)^{-1}\right\|}{\left\|(I+B)^{-1}\right\|}, \frac{1}{\left\|I+B\right\|+L_W}  \right\}
$$
and
\begin{eqnarray*}
\overline{\Delta} &=& \min \Bigg\{ \underline{\theta}, \frac{\overline{M}}{4[(\left\|A\right\|+L_f)K_0+M_g](1+\left\|I+B\right\|+L_W)},\\ && \frac{\overline{M} \left( 1-L_W\left\|(I+B)^{-1}\right\| \right)}{4[(\left\|A\right\|+L_f)K_0+M_g] [1+(1-L_W)\left\|(I+B)^{-1}\right\|]}  \Bigg\}.
\end{eqnarray*}
It is worth noting that the numbers $\epsilon_1$ and $\overline{\Delta}$ are positive according to condition $(A7).$

Suppose that there exists a number $\sigma \in [\zeta, \zeta +\tau]$ such that 
$$
\displaystyle \sup_{t\in[\zeta,\zeta+\tau]} \left\| \phi_{x(t)}(t) - \phi_{\overline{x}(t)}(t) \right\| = \left\| \phi_{x(t)}(\sigma) - \phi_{\overline{x}(t)}(\sigma) \right\|.
$$

Let
$
\kappa=\left\{\begin{array}{ll} \sigma, & ~\textrm{if}~  \sigma \leq \zeta + \tau/2   \\
\sigma - \overline{\Delta}, & ~\textrm{if}~  \sigma > \zeta + \tau/2  \\
\end{array} \right. .\nonumber
$
Since $\overline{\Delta} \le \underline{\theta},$ there exists at most one impulsive moment on the interval $(\kappa,\kappa+\overline{\Delta}).$

First of all, we will consider the case $\sigma > \zeta + \tau/2.$ Assume that there exists an impulsive moment $\theta_{k_0} \in (\kappa, \kappa + \overline{\Delta}).$ For $t \in (\theta_{k_0}, \kappa+ \overline{\Delta}),$ we have that 
\begin{eqnarray*}
&& \left\| \phi_{x(t)}(t) - \phi_{\overline{x}(t)}(t) \right\| \ge  \left\| \phi_{x(t)}(\kappa +\overline{\Delta}) - \phi_{\overline{x}(t)}(\kappa +\overline{\Delta}) \right\|   - \Big\|  \displaystyle \int_{\kappa +\overline{\Delta}}^t  A \left(  \phi_{x(t)}(s) - \phi_{\overline{x}(t)}(s) \right)  ds \Big\|  \\
&& - \Big\|  \displaystyle \int_{\kappa +\overline{\Delta}}^t  \left[ f\left( s,\phi_{x(t)}(s) \right) - f \left( s,\phi_{\overline{x}(t)}(s) \right) \right] ds \Big\|  
 - \Big\|  \displaystyle \int_{\kappa +\overline{\Delta}}^t  [ g(x(s)) - g(\overline{x} (s)) ] ds \Big\|  \\
&& > \overline{M} - 2 \overline{\Delta} [K_0 (\left\|A\right\|+L_f) + M_g]  \\
&& > \displaystyle \frac{\overline{M}}{2} \\
&& \ge \epsilon_1.
\end{eqnarray*}
Making use of the equations 
$$
\phi_{x(t)}(\theta_{k_0}+) = (I+B) \phi_{x(t)}(\theta_{k_0}) + W(\phi_{x(t)}(\theta_{k_0}))
$$
and
$$
\phi_{\overline{x}(t)}(\theta_{k_0}+) = (I+B) \phi_{\overline{x}(t)}(\theta_{k_0}) + W(\phi_{\overline{x}(t)}(\theta_{k_0}))
$$
we obtain that
\[
\left\| \phi_{x(t)}(\theta_{k_0}) - \phi_{\overline{x}(t)}(\theta_{k_0}) \right\| > \displaystyle \frac{\overline{M}-2 \overline{\Delta} [K_0\left(\left\|A\right\|+L_f\right) + M_g] }{\left\|I+B\right\|+L_W}.
\]
By means of the last inequality, one can verify for $t\in (\kappa, \theta_{k_0}]$ that
\begin{eqnarray*}
&& \left\| \phi_{x(t)}(t) - \phi_{\overline{x}(t)}(t) \right\| \ge \left\| \phi_{x(t)}(\theta_{k_0}) - \phi_{\overline{x}(t)}(\theta_{k_0}) \right\| -  \Big\|  \displaystyle \int_{\theta_{k_0}}^t  A \left(  \phi_{x(t)}(s) - \phi_{\overline{x}(t)}(s) \right)  ds \Big\| \\
&& - \Big\|  \displaystyle \int_{\theta_{k_0}}^t  \left[ f\left( s,\phi_{x(t)}(s) \right) - f \left( s,\phi_{\overline{x}(t)}(s) \right) \right] ds \Big\|  
 - \Big\|  \displaystyle \int_{\theta_{k_0}}^t  [ g(x(s)) - g(\overline{x} (s)) ] ds \Big\|  \\
 && > \displaystyle \frac{\overline{M} - 2 \overline{\Delta} [K_0 (\left\|A\right\|+L_f) + M_g] (1+\left\|I+B\right\|+L_W)}{\left\|I+B\right\|+L_W} \\
 && \ge \displaystyle \frac{\overline{M}}{2(\left\|I+B\right\|+L_W)} \\
 && \ge \epsilon_1.
\end{eqnarray*}
Therefore, we have for $t\in (\kappa, \kappa + \overline{\Delta})$ that $\left\| \phi_{x(t)}(t) - \phi_{\overline{x}(t)}(t) \right\| > \epsilon_1.$

On the other hand, if the interval $(\kappa, \kappa + \overline{\Delta})$ does not contain any impulsive moment, then one can confirm that $\left\| \phi_{x(t)}(t) - \phi_{\overline{x}(t)}(t) \right\| > \displaystyle \overline{M} / 2$ for all $t\in (\kappa, \kappa + \overline{\Delta}).$ Hence, the inequality $\left\| \phi_{x(t)}(t) - \phi_{\overline{x}(t)}(t) \right\| > \epsilon_1$ holds for all $t\in (\kappa, \kappa + \overline{\Delta})$ regardless of the existence of an impulsive moment inside the interval.

Next, let us take into account the case $\sigma \le \zeta + \tau/2.$ In the case that the interval $(\kappa, \kappa + \overline{\Delta})$ contains an impulsive moment $\theta_{k_0},$ the inequality
$$  
\left\| \phi_{x(t)}(t) - \phi_{\overline{x}(t)}(t) \right\| > \overline{M} - 2 \overline{\Delta} [K_0 (\left\|A\right\|+L_f) + M_g] > \epsilon_1
$$
is valid for $t \in (\kappa,\theta_{k_0}].$ Therefore, we have that
\begin{eqnarray*}
&& \left\| \phi_{x(t)}(\theta_{k_0}+) - \phi_{\overline{x}(t)}(\theta_{k_0}+) \right\| \ge \left(   \frac{1-L_W \left\|(I+B)^{-1}\right\|}{\left\|(I+B)^{-1}\right\|} \right) \left\| \phi_{x(t)}(\theta_{k_0}) - \phi_{\overline{x}(t)}(\theta_{k_0}) \right\| \\
&& > \left( \frac{1-L_W \left\|(I+B)^{-1}\right\|}{\left\|(I+B)^{-1}\right\|} \right) \left[ \overline{M} - 2 \overline{\Delta}  \left(  K_0 (\left\|A\right\|+L_f) + M_g  \right) \right].
\end{eqnarray*}
The last inequality implies for $t\in (\theta_{k_0}, \kappa+ \overline{\Delta})$ that
\begin{eqnarray*}
&& \left\| \phi_{x(t)}(t) - \phi_{\overline{x}(t)}(t) \right\| > \left( \frac{1-L_W \left\|(I+B)^{-1}\right\|}{\left\|(I+B)^{-1}\right\|} \right)  \overline{M} \\
&& - 2 \overline{\Delta} \left(1+ \frac{1-L_W \left\|(I+B)^{-1}\right\|}{\left\|(I+B)^{-1}\right\|} \right)  [K_0 (\left\|A\right\|+L_f) + M_g] \\
&& \ge \left( \frac{1-L_W \left\|(I+B)^{-1}\right\|}{\left\|(I+B)^{-1}\right\|} \right) \frac{\overline{M}}{2} \\
&& \ge \epsilon_1.
\end{eqnarray*}
If no impulsive moments take place inside the interval $(\kappa, \kappa +\overline{\Delta}),$ then it can be deduced that $$\displaystyle \left\| \phi_{x(t)}(t) - \phi_{\overline{x}(t)}(t) \right\| > \frac{\overline{M}}{2}, \ t\in (\kappa, \kappa +\overline{\Delta}).$$ 
Thus, the inequality $\left\| \phi_{x(t)}(t) - \phi_{\overline{x}(t)}(t) \right\| > \epsilon_1, \ t\in (\kappa, \kappa +\overline{\Delta}),$ is valid for the case $\sigma \le \zeta + \tau/2$ too.  

Now, suppose that there exists an impulsive moment $\theta_{\widetilde{k}} \in [\zeta, \zeta +\tau]$ such that
$$
\displaystyle \sup_{t\in [\zeta, \zeta+\tau]} \left\| \phi_{x(t)}(t) - \phi_{\overline{x}(t)}(t) \right\| = \left\| \phi_{x(t)}(\theta_{\widetilde{k}}+) - \phi_{\overline{x}(t)}(\theta_{\widetilde{k}}+) \right\|.
$$
Let us denote $
\kappa=\left\{\begin{array}{ll} \theta_{\widetilde{k}}, & ~\textrm{if}~  \theta_{\widetilde{k}} \leq \zeta + \tau/2   \\
\theta_{\widetilde{k}} - \overline{\Delta}, & ~\textrm{if}~  \theta_{\widetilde{k}} > \zeta + \tau/2  \\
\end{array} \right. .\nonumber
$

At first, we will consider the case $\theta_{\widetilde{k}} > \zeta + \tau/2.$ 
Since the inequality
$$ \displaystyle
\left\| \phi_{x(t)}(\theta_{\widetilde{k}}) - \phi_{\overline{x}(t)}(\theta_{\widetilde{k}}) \right\| \ge \frac{\left\| \phi_{x(t)}(\theta_{\widetilde{k}}+) - \phi_{\overline{x}(t)}(\theta_{\widetilde{k}}+) \right\|}{\left\|I+B\right\|+L_W}
$$
is valid, one can attain for $t\in (\kappa, \kappa+\overline{\Delta})$ that 
\begin{eqnarray*}
&& \left\| \phi_{x(t)}(t) - \phi_{\overline{x}(t)}(t) \right\| > \frac{\overline{M}}{\left\|I+B\right\|+L_W} - 2 \overline{\Delta} [K_0 (\left\|A\right\|+L_f) + M_g]  \\
&& > \frac{\overline{M}}{2\left(\left\|I+B\right\|+L_W\right)} \\
&& \ge \epsilon_1.
\end{eqnarray*}
On the other hand, if $\theta_{\widetilde{k}} \le \zeta + \tau/2,$ then it can be shown for $t\in (\kappa, \kappa+\overline{\Delta})$ that
\begin{eqnarray*}
\displaystyle \left\| \phi_{x(t)}(t) - \phi_{\overline{x}(t)}(t) \right\| >  \overline{M} - 2 \overline{\Delta} [K_0 (\left\|A\right\|+L_f) + M_g] > \frac{\overline{M}}{2} \ge \epsilon_1.
\end{eqnarray*}
Consequently, system (\ref{impulsive_part}) replicates the sensitivity of (\ref{main_part}). $\square$

\section{Period-doubling cascade} \label{imp_pdc}

In this part of the paper, we suppose that there exists a function $G: \mathbb R \times \mathbb R^m \times \mathbb R \to \mathbb R^m$ satisfying the periodicity condition $G(t+T,x,\mu)=G(t,x,\mu)$ for all $t\in \mathbb R,$ $x\in\mathbb R^m,$ $\mu \in \mathbb R$, where $\mu$ is a parameter, such that for some finite value $\mu_{\infty}$ of the parameter the function $F(t,x)$ on the right-hand side of system (\ref{main_part}) is equal to $G(t,x,\mu_{\infty})$.

System (\ref{main_part}) is said to admit a period-doubling cascade \cite{Feigenbaum80,Sander11,Alligood96,Kovacic11} if there exists a sequence $\left\{\mu_j\right\}$, $j\in \mathbb N$, of period-doubling bifurcation values with $\mu_j \to \mu_{\infty}$ as $j \to \infty$ such that as the parameter $\mu$ increases or decreases through $\mu_j$ the system
\begin{eqnarray} \label{pdc_system}
x'=G(t,x,\mu)
\end{eqnarray}
undergoes a period-doubling bifurcation, i.e., there exists a natural number $\lambda$ such that for each $j\in\mathbb N$ a new periodic solution with period $\lambda 2^j T$ appears in the dynamics of system (\ref{pdc_system}), and consequently, system (\ref{pdc_system}) possesses infinitely many unstable periodic solutions all lying in a bounded region for $\mu=\mu_{\infty}$. 

We say that the impulsive system (\ref{impulsive_part}) replicates the period-doubling cascade of system (\ref{main_part}) if
for each periodic solution $x(t)\in \mathscr{A}$ of (\ref{main_part}) system (\ref{impulsive_part}) admits a periodic solution with the same period.

Under the conditions $(A1)-(A5)$, one can verify using the results of \cite{Akh1,Samolienko95} that if $x(t) \in \mathscr{A}$ is a $\lambda_0 T-$periodic solution of system (\ref{main_part}) for some natural number $\lambda_0$, then the corresponding bounded solution $\phi_{x(t)}(t)$ of (\ref{impulsive_part}) is also $\lambda_0 T-$periodic. Moreover, the instability of all periodic solutions of (\ref{impulsive_part}) is ensured by Theorem \ref{imp_pdc_sensitivity_thm}. Therefore, we have the following theorem.

\begin{theorem} \label{thm_pdc}
If the conditions $(A1)-(A7)$ are valid, then the impulsive system (\ref{impulsive_part}) replicates the period-doubling cascade of system (\ref{main_part}).
\end{theorem}

\begin{remark}
One can confirm that the sequence $\left\{\mu_j\right\}$ of period-doubling bifurcation parameter values is exactly the same for both of the systems (\ref{main_part}) and (\ref{impulsive_part}).  
Therefore, if system (\ref{main_part}) obeys the Feigenbaum universality \cite{Feigenbaum80}, then the same is true also for the impulsive system (\ref{impulsive_part}). 
More precisely, when $\displaystyle \lim_{j \to \infty } \frac{\mu_j-\mu_{j+1}}{\mu_{j+1}-\mu_{j+2}}$ is evaluated, the universal constant known as the Feigenbaum number $4.6692016\ldots$ is achieved, and this universal number is the same for both of the systems (\ref{main_part}) and (\ref{impulsive_part}).
\end{remark}

The next section is devoted to illustrative examples that support the theoretical results.

\section{Examples} \label{sec_examples}

In this part of the paper, two examples will be presented. In the first example the presence of sensitivity in an impulsive system driven by a chaotic Lorenz system will be demonstrated numerically, whereas in the second one period-doubling cascade in an impulsive system driven by a Duffing equation will be discussed.

\textbf{Example 1}

Let us consider the Lorenz system \cite{Lorenz63}
\begin{eqnarray} \label{imppdcex1}
\begin{array}{l}
x'_1=-10x_1+10x_2, \\
x'_2=-x_1 x_3+28 x_1-x_2, \\
x'_3=x_1 x_2-\displaystyle \frac{8}{3} x_3.
\end{array}
\end{eqnarray}
It was demonstrated in \cite{Lorenz63,Sparrow82} that system (\ref{imppdcex1}) is sensitive and it possesses a chaotic attractor.
Next, we take into account the impulsive system
\begin{eqnarray} \label{imppdcex2}
\begin{array}{l}
y'_1=-3 y_1+0.05 \sin(\pi t)+2.4 x_1(t), \\
y'_2=-2 y_2+0.15 \cos y_2-2 x_2(t),  \\
y'_3=-4 y_3+0.2 \tanh y_1+0.6 x_3(t), ~t\neq \theta_k, \\
\Delta y_1 |_{t = \theta_k} = - \displaystyle \frac{2}{3} y_1, \\
\Delta y_2 |_{t = \theta_k} = - \displaystyle \frac{2}{3} y_2 + 0.1 \arctan y_3, \\
\Delta y_3 |_{t = \theta_k} = - \displaystyle \frac{2}{3} y_3,
\end{array}
\end{eqnarray}
where $(x_1(t), x_2(t), x_3(t))$ is a solution of system (\ref{imppdcex1}) and $\theta_k=2k$, $k\in \mathbb Z$. System (\ref{imppdcex2}) is in the form of (\ref{impulsive_part}) with $A=\textrm{diag}(-3,-2,-4),$ $B=\textrm{diag} \left(-\displaystyle \frac{2}{3}, -\displaystyle \frac{2}{3},-\displaystyle \frac{2}{3} \right),$
\[f(t,y_1,y_2,y_3)= \left( 0.05 \sin(\pi t), 0.15 \cos y_2, 0.2 \tanh y_1  \right),\]
\[g(x_1,x_2,x_3)=(2.4 x_1, -2 x_2, 0.6 x_3), ~W(y_1,y_2,y_3)=(0,0.1 \arctan y_3,0).\]

\begin{figure}[ht] 
\centering
\includegraphics[width=8.0cm]{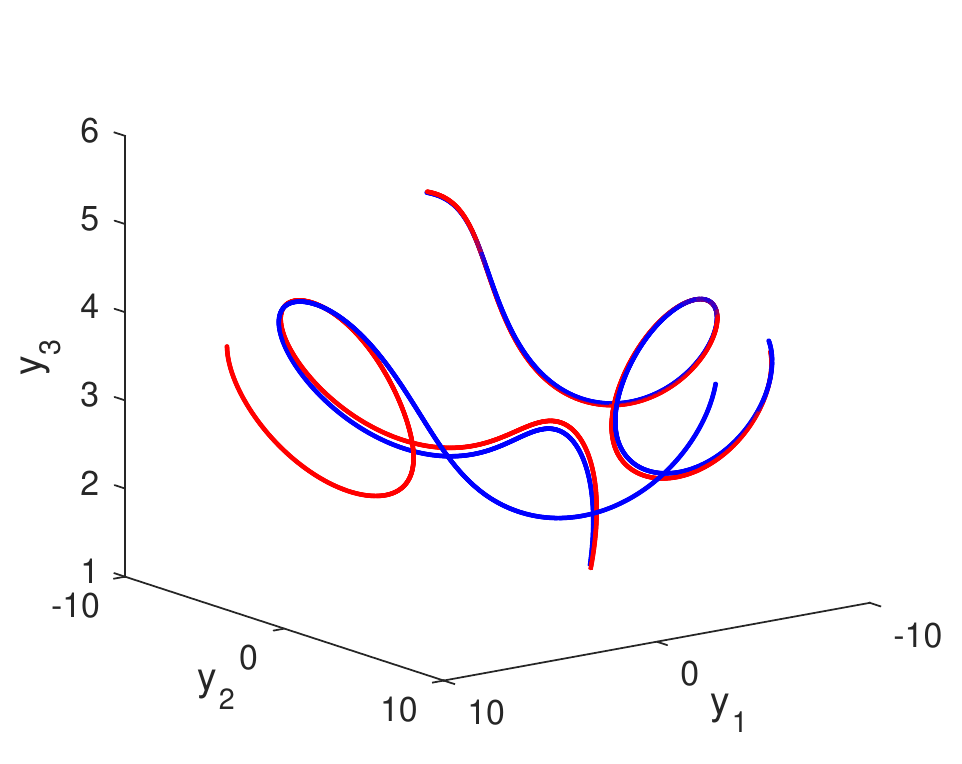}
\caption{Sensitivity in system (\ref{imppdcex2}). The figure manifests the divergence of two initially nearby trajectories shown in red and blue, i.e., the impulsive system (\ref{imppdcex2}) replicates the sensitivity of the Lorenz system (\ref{imppdcex1}). The time interval $[0.5,3.65]$ is used, and both trajectories make a jump at $t=2$.}
\label{impulsivepdcfig1}
\end{figure}

One can verify that the conditions $(A1)-(A7)$ are satisfied for system (\ref{imppdcex2}) with $N=1$, $\omega=2$, $T=2$, $p=1$, $M_f=0.255$, $M_W=0.05 \pi$, $L_f=0.2$, $L_1=0.6$, $L_2=2.4$, and $L_W=0.1$. According to Theorem \ref{imp_pdc_sensitivity_thm}, the impulsive system (\ref{imppdcex2}) replicates the sensitivity of the Lorenz system (\ref{imppdcex1}). Figure \ref{impulsivepdcfig1} shows the $3-$dimensional projections of two initially nearby solutions of the unidirectionally coupled systems (\ref{imppdcex1})-(\ref{imppdcex2}) on the $y_1-y_2-y_3$ space. The trajectory in red corresponds to the initial data $x_1(0.5)=-7.61$, $x_2(0.5)=-2.35$, $x_3(0.5)=33.04$, $y_1(0.5)=-0.53$, $y_2(0.5)=-5.15$, $y_3(0.5)=5.19$, whereas the trajectory in blue corresponds to the initial data $x_1(0.5)=-7.65$, $x_2(0.5)=-2.42$, $x_3(0.5)=33.02$, $y_1(0.5)=-0.51$, $y_2(0.5)=-5.16$, $y_3(0.5)=5.18$. The time interval $[0.5,3.65]$ is used in the simulation, and both trajectories make a jump at $t=2$. Figure \ref{impulsivepdcfig1} supports the result of Theorem \ref{imp_pdc_sensitivity_thm} such that even if the trajectories are initially nearby, later they diverge. In other words, the figure reveals the presence of sensitivity in the impulsive system (\ref{imppdcex2}).

\begin{figure}[ht] 
\centering
\includegraphics[width=8.0cm]{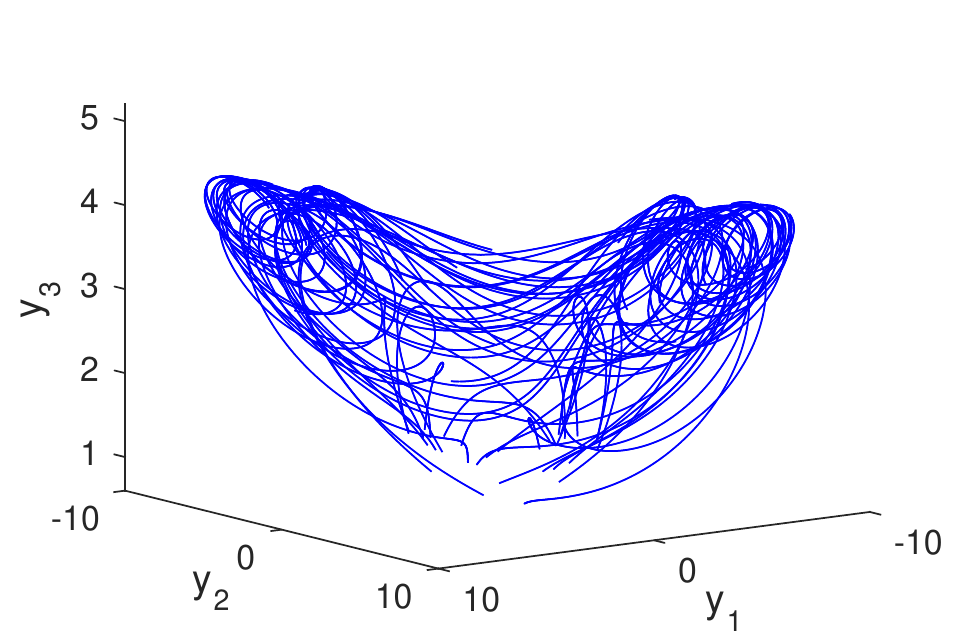}
\caption{Chaotic behavior of system (\ref{imppdcex2}). The discontinuous chaotic trajectory of the impulsive system (\ref{imppdcex2}) supports the result of Theorem \ref{imp_pdc_sensitivity_thm} one more time.}
\label{impulsivepdcfig2}
\end{figure}

In order to show the chaotic behavior of system (\ref{imppdcex2}), we depict in Figure \ref{impulsivepdcfig2} the $3-$dimensional projection of the trajectory of the coupled system (\ref{imppdcex1})-(\ref{imppdcex2}) with $x_1(0.5)=-10.74$, $x_2(0.5)=-13.35$, $x_3(0.5)=26.51$, $y_1(0.5)=-5.94$, $y_2(0.5)=7.67$, $y_3(0.5)=3.52$ on the $y_1-y_2-y_3$ space. The irregular behavior seen in Figure \ref{impulsivepdcfig2} supports the existence of chaos in the dynamics of system (\ref{imppdcex2}). According to the impulse condition in (\ref{imppdcex2}), the chaotic trajectory represented in Figure \ref{impulsivepdcfig2} has discontinuities at the moments $t=\theta_k$. One can confirm that the coupled system (\ref{imppdcex1})-(\ref{imppdcex2}) is also chaotic.

\textbf{Example 2}

It was demonstrated in paper \cite{Sato83} that the Duffing equation
\begin{eqnarray} \label{example2_1}
x''+0.3x'+x^3=\mu \cos t, 
\end{eqnarray}
where $\mu$ is a parameter, displays period-doubling bifurcations and leads to chaos at $\mu=\mu_{\infty}\equiv 40$. 

Using the variables $x_1=x$ and $x_2=x',$ equation (\ref{example2_1}) can be rewritten as a system in the form
\begin{eqnarray} 
\begin{array}{l}
x'_1=x_2, \\ \label{example2_2}
x'_2=-0.3x_2-x_1^3+\mu \cos t. 
\end{array}
\end{eqnarray}
One can confirm that the chaotic attractor of system (\ref{example2_2}) takes place inside the compact region
\begin{eqnarray*}
\Lambda= \left\{(x_1,x_2) \in \mathbb R^2 :~ \left|x_1\right| \leq 5.5, ~ \left|x_2\right| \leq 14 \right\}.
\end{eqnarray*}

Next, we consider the impulsive system
\begin{eqnarray} \label{example2_3}
\begin{array}{l}
y'_1=- y_1-4y_2 +0.12 \arctan y_2 + 2.1 x_1(t)-0.3 \sin (x_1(t)), \\
y'_2= y_1-3y_2 +0.7\cos t-1.6 x_2(t)+0.01x_2^2(t), ~t\neq \theta_k, \\
\Delta y_1 |_{t = \theta_k} = - \displaystyle \frac{1}{2} y_1 +0.08\sin y_2, \\
\Delta y_2 |_{t = \theta_k} = - \displaystyle \frac{1}{2} y_2,
\end{array}
\end{eqnarray}
where $(x_1(t),x_2(t))$ is a solution of (\ref{example2_2}) and $\theta_k= \pi k,$ $k\in \mathbb Z$.

The impulsive system (\ref{example2_3}) is in the form of (\ref{impulsive_part}) with
$A=\left(
\begin {array}{ccc}
-1&-4\\
\noalign{\medskip}
1&-3
\end {array}
\right),$ $B=\textrm{diag} \displaystyle \left(-\frac{1}{2}, -\frac{1}{2} \right),$ 
$f(t,y_1,y_2)= \left( 0.5 \arctan y_2, 0.7 \cos t \right),$
$g(x_1,x_2)=(2.1 x_1-0.3\sin x_1, -1.6 x_2+ 0.01 x^2_2),$ and $W(y_1,y_2)=(0.08 \sin y_2, 0).$

Let us denote by $U(t,s)$ the transition matrix of the linear homogeneous system
\begin{eqnarray} \label{example2_4}
\begin{array}{l}
u'_1=- u_1-4u_2, \\
u'_2= u_1-3u_2, ~t\neq \theta_k, \\
\Delta u_1 |_{t = \theta_k} = - \displaystyle \frac{1}{2} u_1, \\
\Delta u_2 |_{t = \theta_k} = - \displaystyle \frac{1}{2} u_2.
\end{array}
\end{eqnarray}
It can be verified that
$$
U(t,s)=e^{-2(t-s)} \displaystyle \left( \frac{1}{2}\right)^{i([s,t))} P 
\left(
\begin {array}{ccc}
\cos(\sqrt{3}(t-s))&-\sin(\sqrt{3}(t-s))\\
\noalign{\medskip}
\sin(\sqrt{3}(t-s))&\cos(\sqrt{3}(t-s))
\end {array}
\right) P^{-1}, ~t > s,
$$
where $P=\left(
\begin {array}{ccc}
\sqrt{3}&1\\
\noalign{\medskip}
0&1
\end {array}
\right).$
The conditions $(A1)-(A7)$ are satisfied for system (\ref{example2_3}) with $N=2.48421$, $\omega=2 $, $T=2\pi$, $p=2$, $M_f=0.72494$, $M_W= 0.08$, $L_f= 0.12$, $L_1= 1.32$, $L_2=2.4$, and $L_W=0.08$.
Moreover, the eigenvalues of the matrix 
$$A+\displaystyle \frac{p}{T} \ln(I+B)=\left(
\begin {array}{ccc}
-1-\displaystyle \frac{\ln2}{\pi}&-4\\
\noalign{\medskip}
1&-3-\displaystyle \frac{\ln2}{\pi}
\end {array}
\right)$$
are $-2- \displaystyle \frac{\ln 2}{\pi} \pm i \sqrt{3}$.

\begin{figure}[ht] 
\centering
\includegraphics[width=14.0cm]{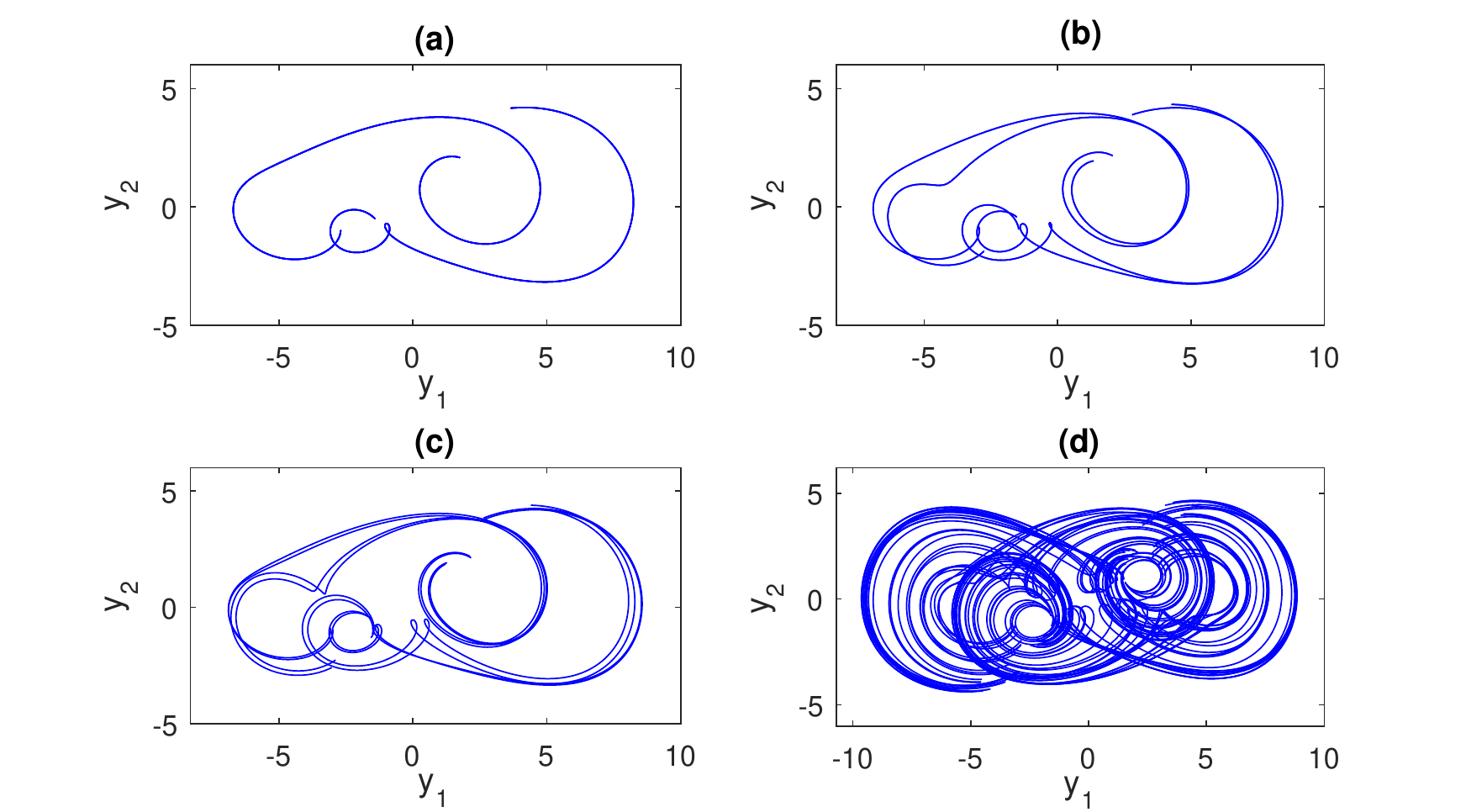}
\caption{Periodic and chaotic orbits of the impulsive system (\ref{example2_3}). (a) Period$-1$ orbit. (b) Period$-2$ orbit. (c) Period$-4$ orbit. (d) Chaotic orbit. The parameter values $\mu=31.7$, $\mu=34.3$, $\mu=36.1$, and $\mu=40$ are utilized in (a), (b), (c), and (d), respectively. The figure shows the presence of a period-doubling cascade in the dynamics of (\ref{example2_3}) such that the system leads to chaos at $\mu=40$.}
\label{impulsivepdcfig3}
\end{figure}

\begin{figure}[ht] 
\centering
\includegraphics[width=13.0cm]{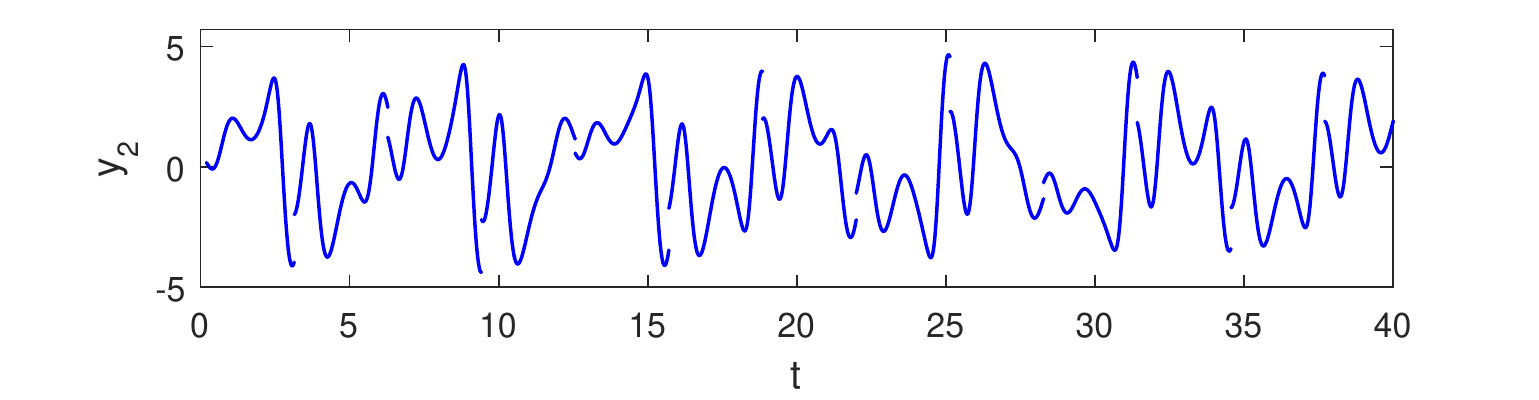}
\caption{Time-series of the $y_2-$coordinate of system (\ref{example2_3}) with $\mu=40$. The initial data $x_1(0.2)=3.16$, $x_2(0.2)=1.86$, $y_1(0.2)=0.71$, and $y_2(0.2)=0.18$ are utilized. The figure reveals the presence of chaos in the dynamics of system  (\ref{example2_3}).}
\label{impulsivepdcfig4}
\end{figure}

According to Theorem \ref{thm_pdc}, the impulsive system (\ref{example2_3}) replicates the period-doubling cascade of  the Duffing equation (\ref{example2_1}) and possesses chaos at the parameter value $\mu=40$.
In order to demonstrate the period-doubling cascade of (\ref{example2_3}) numerically, we depict in Figure \ref{impulsivepdcfig3} the periodic orbits as well as a chaotic trajectory. Figure \ref{impulsivepdcfig3}, (a), (b), and (c) respectively show the period$-1$, period$-2$, and period$-4$ orbits of (\ref{example2_3}). The parameter values $\mu=31.7$, $\mu=34.3$, and $\mu=36.1$ are utilized in Figure \ref{impulsivepdcfig3}, (a), (b), and (c), respectively. On the other hand, taking $\mu=40$ in the coupled system (\ref{example2_2})-(\ref{example2_3}), we represent in Figure \ref{impulsivepdcfig3} (d) the $2-$dimensional projection of the trajectory of (\ref{example2_2})-(\ref{example2_3}) corresponding to the initial data $x_1(0.2)=3.16$, $x_2(0.2)=1.86$, $y_1(0.2)=0.71$, $y_2(0.2)=0.18$ on the $y_1-y_2$ plane. Figure \ref{impulsivepdcfig3} supports the result of Theorem \ref{thm_pdc} such that the impulsive system (\ref{example2_3}) admits chaos through period-doubling cascade at the parameter value $\mu=40$. Moreover, we represent in Figure \ref{impulsivepdcfig4}, the time-series of the $y_2-$coordinate of the trajectory shown in Figure \ref{impulsivepdcfig3}, (d). The irregular behavior of the time-series also supports the presence of chaos in system (\ref{example2_3}).

\end{document}